\documentclass[10pt,conference]{IEEEtran}
\IEEEoverridecommandlockouts
\usepackage{cite}
\usepackage{amsmath,amssymb,amsfonts}
\usepackage{algorithm}
\usepackage{algpseudocode}
\usepackage{graphicx}
\usepackage{textcomp}
\usepackage{xcolor}
\usepackage{balance}
\usepackage{tikz}
\usepackage{pgfplots}
\pgfplotsset{compat=1.6}
\def\BibTeX{{\rm B\kern-.05em{\sc i\kern-.025em b}\kern-.08em
    T\kern-.1667em\lower.7ex\hbox{E}\kern-.125emX}}
\begin{document}
\bstctlcite{IEEEexample:BSTcontrol}
\title{countBF: A General-purpose High Accuracy and Space Efficient Counting Bloom Filter}

\author{\IEEEauthorblockN{1\textsuperscript{st} Sabuzima Nayak}
\IEEEauthorblockA{\textit{Dept. Computer Science \& Engineering} \\
\textit{National Institute of Technology Silchar}\\
Cachar-788010, Assam, India\\
sabuzima\_rs@cse.nits.ac.in}
\and
\IEEEauthorblockN{2\textsuperscript{nd} Ripon Patgiri,~\IEEEmembership{Senior Member, IEEE}}
\IEEEauthorblockA{\textit{Dept. Computer Science \& Engineering} \\
\textit{National Institute of Technology Silchar}\\
Cachar-788010, Assam, India\\
ripon@cse.nits.ac.in}}

\maketitle

\begin{abstract}
Bloom Filter is a probabilistic data structure for the membership query, and it has been intensely experimented in various fields to reduce memory consumption and enhance a system's performance. Bloom Filter is classified into two key categories: counting Bloom Filter (CBF), and non-counting Bloom Filter. CBF has a higher false positive probability than standard Bloom Filter (SBF), i.e., CBF uses a higher memory footprint than SBF. But CBF can address the issue of the false negative probability. Notably, SBF is also false negative free, but it cannot support delete operations like CBF. To address these issues, we present a novel counting Bloom Filter based on SBF and 2D Bloom Filter, called countBF.  countBF uses a modified murmur hash function to enhance its various requirements, which is experimentally evaluated. Our experimental results show that countBF uses $1.96\times$ and $7.85\times$ less memory than SBF and CBF respectively, while preserving lower false positive probability and execution time than both SBF and CBF. The overall accuracy of countBF is $99.999921$, and it proves the superiority of countBF over SBF and CBF. Also, we compare with other state-of-the-art counting Bloom Filters.
\end{abstract}

\begin{IEEEkeywords}
Bloom Filter, Counting Bloom Filter, Membership Filter, Frequency Count, Count-Min Sketch, Data Structures.
\end{IEEEkeywords}

\section{Introduction}
Bloom Filter \cite{Bloom} is an extensively experimented data structure. It has met a vast application domains, namely, Big Data \cite{Fuzzy}, IoT \cite{Singh}, Computer Networking \cite{Rabieh}, Network Security and Privacy \cite{DDoS,PassDB}, Biometrics \cite{Gomez}, and Bioinformatics \cite{Nayak}. Counting Bloom Filter is useful in Computer Network, Network Security and Privacy \cite{Hunt}. Therefore, there are diverse Bloom Filters available to address the various problems of diverse domains \cite{Dorojevets,Var,Kiss,cmin,Reviriego}. Bloom Filter is categorized into two key categories, namely, counting and non-counting Bloom Filter. Non-counting Bloom Filter (conventional) is faster than counting Bloom Filter. However, the non-counting Bloom Filter does not support delete operation due to a false negative issue. Delete operation introduces a false positive issue. On the contrary, counting Bloom Filter supports delete operation and can solve false negative issue \cite{countingBF}. However, the false positive probability counting Bloom Filter is higher than the conventional Bloom Filter \cite{Kirsch,KM}. Alternatively, Bloom Filter occupies more memory to achieve the desired false positive probability than conventional Bloom Filter. 

Delete operation is crucial for Bloom Filter.  Suppose a database management system integrates Bloom Filter to avoid unnecessary disk accesses and enhance its performance significantly with a tiny amount of memory \cite{BigTable}. Thus, the database management system requires to insert, query, and delete operation for which conventional Bloom Filter is not suitable. Therefore, counting Bloom Filter is used in such kind of requirements. Another example is a network router. Older packets are removed from the router databases. Therefore, the router requires counting Bloom Filter. Moreover, counting Bloom Filter is an effective approximation tool for frequency count. Therefore, counting Bloom Filter is adapted in diverse applications. Moreover, there are numerous variants of counting Bloom Filters available. Rottenstreich \textit{et al.} \cite{Rottenstreich} develops an excellent counting Bloom Filter based on $B_h$ sequences, called $B_h$-CBF and variable-increment counting Bloom Filter (VI-CBF). Unlike conventional counting Bloom Filter, insertion in $B_h$-CBF is incremented based on $B_h$ sequences. $B_h$-CBF able to reduce false positives, however, the space consumption is very high. In addition, Pontarelli \textit{et al.} \cite{Pontarelli} develops a new counting Bloom Filter based on fingerprint, called FP-CBF. FP-CBF uses fingerprints to improve the counting Bloom Filter, and therefore, it requires $(C+F)\times m$, where $C$ is the number of bits, $F$ is the fingerprint size, and $m$ is the total number of positions. FP-CBF uses $(k+1)$ hash functions to insert, delete and lookup operations. Also, FP-CBF outperforms VI-CBF in terms of false positive and space consumption. Moreover, Ternary Bloom Filter (TBF) \cite{Lim} is a counting Bloom Filter to address the false positive and false negative.

As we know that counting Bloom Filter has a high false positive probability for which it requires a higher memory footprint than conventional Bloom Filter. CBF can eradicate the false negative issues, but it has a high false positive probability, and high memory footprint. To lower the false positive probability, counting Bloom Filter sacrifices memory footprint. Therefore, we propose a novel counting Bloom Filter to address the above-raised issues, called countBF. The countBF can reduce the false positive probability significantly while preserving a low memory footprint. Our experimental results show that countBF outperforms standard Bloom Filter (SBF) \cite{Kirsch} and counting Bloom Filter \cite{countingBF} in every aspect. Key objectives of our proposed system is to reduce memory footprint, to lower false positive probability and to increase its accuracy without compromising the insertion/query performance.

Our proposed counting Bloom Filter is similar to the conventional Bloom Filters. countBF has counters, while SBF does not have counters. Also, countBF is implemented in the platform of a 2D Bloom Filter (2DBF) \cite{rDBF}. 2DBF uses a 2D integer array instead of relying on the bitmap array. This 2D integer array is used as a bitmap where each integer represents a block of bits. countBF enhances its performance by tuning the murmur hash function \cite{murmur}. Murmur hash function is the best non-cryptographic string hash functions \cite{stat}. There are also cryptographic string hash functions, however, it does not enhances the performance and the false positive probability \cite{ICOIN}. Therefore, we compare countBF with SBF and CBF to evaluate the characteristics using various test cases. In our experimental work, we have compared countBF with SBF because counting Bloom Filter is unable outperform SBF, and thus, it is justified to compare with SBF and CBF. Also, we compare with the other filters with countBF.


\section{countBF: The Proposed System}
\label{PS}
\begin{figure}[!ht]
    \centering
    \includegraphics[width=0.4\textwidth]{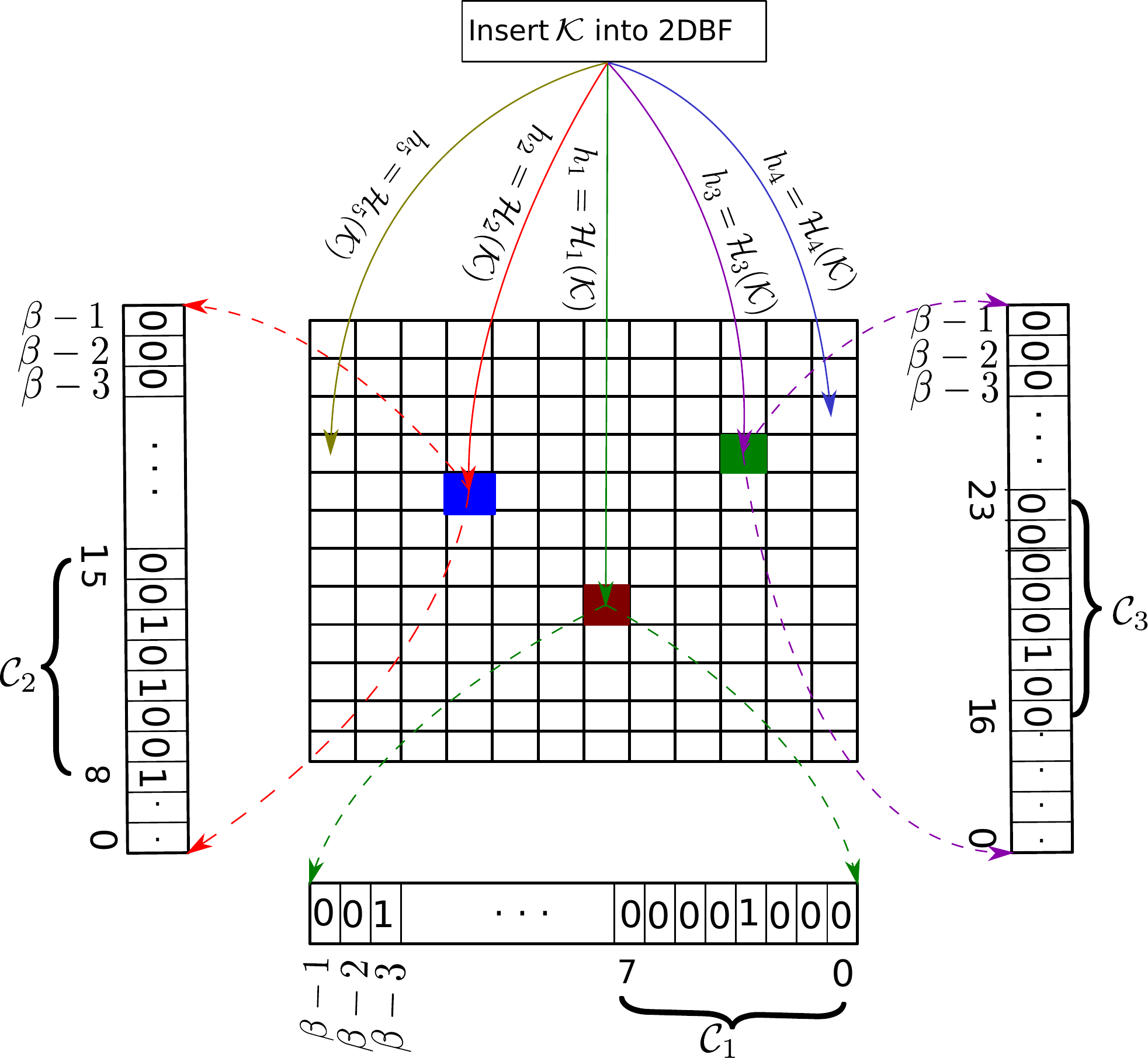}
    \caption{Architecture of countBF with 8 bit counters}
    \label{arc}
\end{figure}
We present a novel counting Bloom Filter, called countBF, by deploying 2-Dimensional Bloom Filter \cite{rDBF}. countBF uses a few arithmetic operations to increase its performance. Let, $\mathbb{B}_{x,y}$ be the two-dimensional integer array to implement counting Bloom Filter where $x$ and $y$ are the dimensions of the filter. The $x\not=y$ and these are prime numbers. A cell of the $\mathbb{B}_{x,y}$ is constituted by $\eta$ counters, as shown in Figure \ref{arc}. We have demonstrated 8 bits counters for an example in Figure \ref{arc}. Each counter contains $\alpha$ bits, which is defined by the user, and it can be 1 to 64 bits depending on the user's requirement. The data type of a cell is \textbf{unsigned long int} or \textbf{unsigned int}. Each cell of $\mathbb{B}_{x,y}$ occupies $\beta~bits$. Therefore, the total number of counters is calculated as $\eta=\frac{\beta}{\alpha}$. Particularly, $\eta=8$ counters, if each cell occupies $\beta=64$ bits and each counter contains $\alpha=8$ bits. Our key objective is to reduce memory footprint, lower the false positive probability, increase the accuracy and enhance the query performance of the counting Bloom Filter. countBF uses two masks, namely, extract mask and reset mask. An extract mask is used to extract the bit information of a counter from a cell. Similarly, a reset mask is used to reset the bit information of a counter in a cell. These two masks are used because of countBF relies on the integer array instead of the bitmap array.

The extract mask are dependent on the number of counters $\eta$. Therefore, there are $\eta$ extract masks which are stored in an array. Extract mask is defined as $\mathcal{M}^e=\{\mathcal{M}^e_1,\mathcal{M}^e_2,\mathcal{M}^e_3,\ldots,\mathcal{M}^e_\eta\}$. For instance, $\mathcal{M}^e_2=\ldots 000001111111100000000$ for 8 bits counters. The third extract mask is $0x00000000001FC000$ and it is the correct representation of the extract mask in 64 bits measures for 7 bits counter. Extract masks are used to extract certain counter's value using bit operations. For instance, \textbf{unsigned long int} occupies $64$ bits and there are 8 counters with 8 bits each. To extract $l^{th}$ counter information, we perform $\mathcal{C}_l=(\mathbb{B}_{i,j}~\land~\mathcal{M}^e_l)$. To remove the trailing zeros, we perform $\mathcal{C}_l=(\mathcal{C}_l>>(\eta*l))$. This $\mathcal{C}_l$ gives the counter information. 

Similar to extract mask, there are also $\eta$ masks for reset a counter to zero. Reset mask is defined as $\mathcal{M}^r=\{\mathcal{M}^r_1,\mathcal{M}^r_2,\mathcal{M}^r_3,\ldots,\mathcal{M}^r_\eta\}$. For instance, $\mathcal{M}^r_2=\ldots 111110000000011111111$ for a 8 bit counter. The third reset mask is $0xFFFFFFFFFFE03FFF$ and it is the correct representation of the reset mask in 64 bits measures for 7 bits counter. Extract masks are used to extract certain counter's value. Reset masks are used to reset the counter's value to zero. To reset $l^{th}$ counter's value to zero, we need to perform $\mathbb{B}_{i,j}=(\mathbb{B}_{i,j}~\land~\mathcal{M}^r_l)$.

The counting Bloom Filter is comprised of a set of counters to counts the input items. Conventional Bloom Filter uses bitmap array to manipulate the bits to store information of input items. However, countBF does not use bitmap arrays. Instead, it uses a 2D integer array where each cell occupies some memory depending on the data type. For instance, \textbf{unsigned long int} occupies 64 bits in modern computers, and therefore, each cell occupies 64 bits memory, and it is initialized by zero. Let $\eta$ be the total number of counters, $\alpha$ be the bits per counters, and $\mu$ be the bits per cell in a 2D array. Therefore, the total number of counters in each cell of countBF is $\eta=\frac{\mu}{\alpha}$, and the remainder is not used. Thus, the total number of masks varies depending on the bits used per counter $\alpha$.


\begin{algorithm}
\caption{Insertion of an item $\mathcal{K}$ into countBF using $k$ hash functions.}
\begin{algorithmic}[1]
\Procedure{Insertion}{$\mathbb{B}_{x,y},~\mathcal{K}$}
    \For {$i=1~to~k$}
        \State $h=\mathcal{H}_i(\mathcal{K},Seed_i)$
        \State $\Call{Increment}{\mathbb{B}_{x,y},h}$
    \EndFor
\EndProcedure
\end{algorithmic}
\label{Algo1}
\end{algorithm}

\begin{algorithm}
\caption{Increment a single counter while inserting an item $\mathcal{K}$ into countBF using a single hash function.}
\begin{algorithmic}[1]
\Procedure{Increment}{$\mathbb{B}_{x,y},~h$}
    \State $i=h\%~x,~j=h\%~y,~l=h\%~\eta$
    \State $\mathcal{C}_l=\mathbb{B}_{i,j}~\land~\mathcal{M}^e_l$
    \State $\mathcal{C}_l=\mathcal{C}_l>>(\alpha*l)$
    \State $\mathcal{C}_l=\mathcal{C}_l+1$
    \If{$\mathcal{C}_l=MAX$}
        \State Counter Overflow.
        \State $\Return$
    \EndIf
    \State $\mathcal{C}_l=\mathcal{C}_l<<(\alpha*l)$
    \State $\mathbb{B}_{i,j}=\mathbb{B}_{i,j}~\land~\mathcal{M}^r_l$
    \State $\mathbb{B}_{i,j}=\mathbb{B}_{i,j}~\lor~\mathcal{C}_l$
\EndProcedure
\end{algorithmic}
\label{Algo2}
\end{algorithm}

countBF is a counting Bloom Filter that comprises many counters. The counters are incremented upon insertion of an item. Let, $\mathbb{B}_{i,j}$ be a 2D Bloom Filter (2DBF) and $\mathcal{C}_l$ be the $l^{th}$ counter in a cell of a 2DBF. Let, $\mathcal{H}()$ be a hash function. We use the murmur hash function. Difference seed values create a different hash value for the same key. For insertion of a single item, countBF calls $k$ hash functions, and the item is inserted into the $k$ counters. Algorithm \ref{Algo1} demonstrates insertion of an item $\mathcal{K}$ using $k$ hash functions. It requires increment the counters' value. Algorithm \ref{Algo2} shows the incrementing process of a counter.

\begin{algorithm}
\caption{Lookup an item $\mathcal{K}$ in countBF using $k$ hash functions.}
\begin{algorithmic}[1]
\Procedure{Lookup}{$\mathbb{B}_{x,y},~\mathcal{K}$}
    \For {$i=1~to~k$}
        \State $h=\mathcal{H}_i(\mathcal{K},Seed_i)$
        \State $flag\leftarrow flag~\land~\Call{Test}{\mathbb{B}_{x,y},h}$
    \EndFor
    \State $\Return~flag$
\EndProcedure
\end{algorithmic}
\label{Algo3}
\end{algorithm}

\begin{algorithm}
\caption{Lookup in a single counter while query an item $\mathcal{K}$ in countBF using a single hash function.}
\begin{algorithmic}[1]
\Procedure{Test}{$\mathbb{B}_{x,y},Hashvalue~h$}
    \State $i=h\%~x,~j=h\%~y,~l=h\%~\eta$
    \State $\mathcal{C}_l=\mathbb{B}_{i,j}~\land~\mathcal{M}^e_l$
    \State $\mathcal{C}_l=\mathcal{C}_l>>(\alpha*l)$
    \If{$\mathcal{C}_l\ge 1$}
        \State $\Return~true$
    \Else
        \State $\Return~false$
    \EndIf
\EndProcedure
\end{algorithmic}
\label{Algo4}
\end{algorithm}

Lookup or query operation is similar to insertion operation except the increment steps. Querying an item $\mathcal{K}$ requires $k$ hash functions with $k$ seed values. The seed values and hash functions of lookup procedure cannot be different from the seed values and hash functions of insertion and delete operations. Algorithm \ref{Algo3} calls murmur hash function $k$ times and \textsc{Test}() function $k$ times. The \textsc{Test}() function is demonstrated in Algorithm \ref{Algo4} that returns either $true$ or $false$. The variable $flag$ holds the final result of all test functions and returns the variable $flag$. The $flag$ is a Boolean variable that can hold either $true$ or $false$. All \textsc{Test}() function results are $AND$ed which produce Boolean value of $true$ or $false$ and assigned to $flag$.

\begin{algorithm}
\caption{Delete an item $\mathcal{K}$ in countBF using $k$ hash functions.}
\begin{algorithmic}[1]
\Procedure{Delete}{$\mathbb{B}_{x,y},Keys~\mathcal{K}$}
    \For {$i=1~to~k$}
        \State $h=\mathcal{H}_i(\mathcal{K},Seed_i)$
        \State $\Call{Decrement}{\mathbb{B}_{x,y},h}$
    \EndFor
\EndProcedure
\end{algorithmic}
\label{Algo5}
\end{algorithm}

\begin{algorithm}
\caption{Decrement a single counter while deleting an item $\mathcal{K}$ from countBF using a single hash function.}
\begin{algorithmic}[1]
\Procedure{Decrement}{$\mathbb{B}_{x,y},Hashvalue~h$}
    \State $i=h\%~x,~j=h\%~y,~l=h\%~\eta$
    \State $\mathcal{C}_l=\mathbb{B}_{i,j}~\land~\mathcal{M}^e_l$
    \State $\mathcal{C}_l=\mathcal{C}_l>>(\alpha*l)$
    \State $\mathcal{C}_l=\mathcal{C}_l-1$
    \If{$\mathcal{C}_l<1$}
        \State No deletion.
        \State $\Return$
    \EndIf
    \State $\mathcal{C}_l=\mathcal{C}_l<<(\alpha*l)$
    \State $\mathbb{B}_{i,j}=\mathbb{B}_{i,j}~\land~\mathcal{M}^r_l$
    \State $\mathbb{B}_{i,j}=\mathbb{B}_{i,j}~\lor~\mathcal{C}_l$
\EndProcedure
\end{algorithmic}
\label{Algo6}
\end{algorithm}

Conventional Bloom Filter does not support delete operation due to false negatives. The counting Bloom Filter was introduced to address the issue of false negatives \cite{countingBF}. The delete operation creates the issue of false negatives in conventional Bloom Filter. Therefore, counting Bloom Filter is used in many domains. Similar to conventional counting Bloom Filter, countBF also supports a delete operation without any false negative issue. To delete an item $\mathcal{K}$, countBF requires $k$ hash functions call and $k$ \textsc{Decrement}() functions calls as demonstrated in Algorithm \ref{Algo5}. The insertion and delete operations are required the same steps except for the decrement of the counters in delete operation, as shown in Algorithm \ref{Algo6}.

\section{Experimental Results}
\label{ER}
Our proposed algorithm is evaluated in 8GB RAM, Intel® Core™ i7-7700 CPU @ 3.60GHz $\times$ 8, Ubuntu 18.04.5 LTS and GCC version 7.5.0. We created four different datasets, namely, Same Set, Mixed Set, Disjoint Set, and Random Set. Let, $\mathcal{S}=\{x_1,x_2,x_3,\ldots,x_n\}$ be an inserted set into the Bloom Filter, $\mathcal{Q}$ be the query set. The Same Set defines $\mathcal{S}=\mathcal{Q}$ whereas Disjoint Set is defined as $\mathcal{S}\cap\mathcal{Q}=\phi$. The definition of the Mixed Set follows any one condition, either $q_1\in\mathcal{S}$ and $q_2\not\in\mathcal{S}$ or $q_1\not\in\mathcal{S}$ and $q_2\in\mathcal{S}$ where $q_1\subset\mathcal{Q}$ and $q_2\subset\mathcal{Q}$. However, the Random Set is randomly generated dataset. These test cases are able to unearth the strengths and weaknesses of a Bloom Filter. We have assessed our proposed countBF for various counter's sizes for the fair judgement.

\pgfplotstableread[row sep=\\,col sep=&]{
interval& three&	four&	five&	six&	seven&	eight&  sbf&  cbf\\
10M&	1.552348&	1.542702&	1.558794&	1.604979&	1.604359&	1.544543&	3.97461&	5.199482\\
20M&	3.215669&	3.207123&	3.208421&	3.228241&	3.144407&	3.138284&	9.120941&	10.78577\\
30M&	4.831773&	4.837797&	4.816114&	4.753442&	4.751931&	4.735329&	13.761192&	17.236443\\
40M&	6.426489&	6.412976&	6.43806&	6.367437&	6.360259&	6.430788&	18.236497&	23.229478\\
50M&	8.024412&	8.27945&	8.02449&	7.968762&	7.960264&	8.130709&	23.183817&	28.910923\\
}\ins
\begin{figure}[!ht]
\centering
\begin{tikzpicture}
    \begin{axis}[
            ybar,
            bar width=.1cm,
            width=0.48\textwidth,
            height=.2\textwidth,
            enlarge x limits=0.1,
            legend style={at={(0.5,1)},
                anchor=south,legend columns=4,legend cell align=left},
            symbolic x coords={10M,20M,30M,40M,50M},
            xtick=data,
             x tick label style={rotate=45,anchor=east},
            nodes near coords align={vertical},
            ymin=1,ymax=32,
            ymode=log,
            ylabel={Time in Second},
        ]
        \addplot table[x=interval,y=three]{\ins};
        \addplot table[x=interval,y=four]{\ins};
        \addplot table[x=interval,y=five]{\ins};
        \addplot table[x=interval,y=six]{\ins};
        \addplot table[x=interval,y=seven]{\ins};
        \addplot table[x=interval,y=eight]{\ins};
        \addplot table[x=interval,y=sbf]{\ins};
        \addplot table[x=interval,y=cbf]{\ins};
        \legend{3 bits,4 bits, 5 bits, 6 bits, 7 bits, 8 bits, SBF, CBF}
    \end{axis}
\end{tikzpicture}
\caption{Comparison of insertion time of countBF using 3 bits, 4 bits, 5 bits, 6 bits, 7 bits, and 8 bits counters with SBF and CBF. Lower is better.}
\label{ins}
\end{figure}
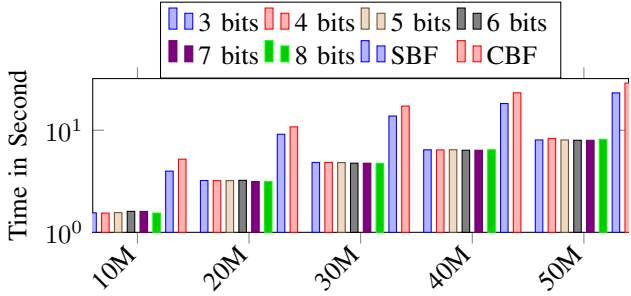

countBF is compared with standard Bloom Filter (SBF) \cite{Kirsch} and conventional counting Bloom Filter (CBF) \cite{countingBF}. The countBF is evaluated by setting the counter's bits by 3 bits, 4 bits, 5 bits, 6 bits, 7 bits, and 8 bits. A counter's bit can be a maximum of 64-bits. In our experimental evaluation, each cell occupies 64 bits, and hence, each cell has different counters. The total number of counters for 3 bits, 4 bits, 5 bits, 6 bits, 7 bits, and 8 bits counters are 21, 16, 12, 10, 9, and 8 counters in each cell of countBF. In this configuration, we conduct the experiments, and Figure \ref{ins} demonstrates the insertion time taken by countBF, SBF, and CBF. On an average, countBF is faster than SBF and CBF in the insertion of items. countBF with 4 bits counter is slowest among countBF with  3 bits, 5 bits, 6 bits, 7 bits, and 8 bits counters and countBF with 7 bits counter is fastest among countBF with 3 bits, 4 bits, 5 bits, 6 bits, 7 bits, and 8 bits counters in the insertion of items. Individually, countBF with 4 bits, 8 bits, 7 bits, 7 bits, and 7 bits counters are the fastest in the insertion of 10M, 20M, 30M, 40M, and 50M dataset, respectively. Overall, countBF with 7 bits is the fastest counter in the insertion operation.

\pgfplotstableread[row sep=\\,col sep=&]{
interval&  three&	four&	five&	six&	seven&	eight&  sbf&  cbf\\
Same Set&	1.533347&	1.540859&	1.578427&	1.52912&	1.526925&	1.54442&	3.875316&	4.56835\\
Mixed Set&	1.357777&	1.357747&	1.359726&	1.365289&	1.364774&	1.495714&	3.159874&	3.513943\\
Disjoint Set&	1.471781&	1.468922&	1.478426&	1.469426&	1.479686&	1.679743&	2.559842&	2.758831\\
Random Set&	1.649408&	1.655802&	1.657206&	1.652414&	1.646859&	1.88201&	2.606786&	2.723711\\
}\lo
\begin{figure}[!ht]
\centering
\begin{tikzpicture}
    \begin{axis}[
            ybar,
            bar width=.1cm,
            width=0.48\textwidth,
            height=.2\textwidth,
            enlarge x limits=0.1,
            legend style={at={(0.5,1)},
                anchor=south,legend columns=4,legend cell align=left},
            symbolic x coords={Same Set, Mixed Set, Disjoint Set, Random Set},
            xtick=data,
             x tick label style={rotate=45,anchor=east},
            nodes near coords align={vertical},
            ymin=1,ymax=5,
            ymode=log,
            ylabel={Time in Second},
        ]
        \addplot table[x=interval,y=three]{\lo};
        \addplot table[x=interval,y=four]{\lo};
        \addplot table[x=interval,y=five]{\lo};
        \addplot table[x=interval,y=six]{\lo};
        \addplot table[x=interval,y=seven]{\lo};
        \addplot table[x=interval,y=eight]{\lo};
        \addplot table[x=interval,y=sbf]{\lo};
        \addplot table[x=interval,y=cbf]{\lo};
        \legend{3 bits,4 bits, 5 bits, 6 bits, 7 bits, 8 bits, SBF, CBF}
    \end{axis}
\end{tikzpicture}
\caption{Comparison of 10M items lookup time by countBF using 3 bits, 4 bits, 5 bits, 6 bits, 7 bits, and 8 bits counters with SBF and CBF. Lower is better.}
\label{look1}
\end{figure}
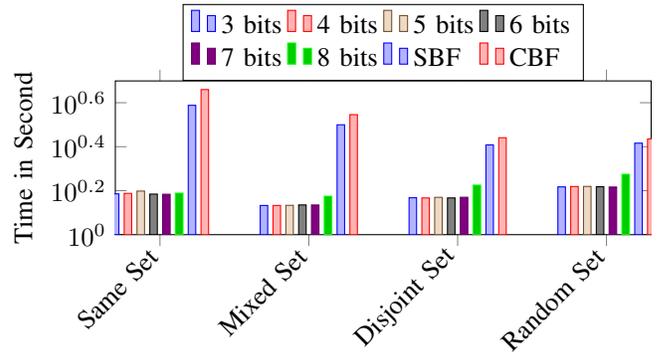

\pgfplotstableread[row sep=\\,col sep=&]{
interval&  three&	four&	five&	six&	seven&	eight&   sbf&  cbf\\
Same Set&	3.122606&	3.115545&	3.111003&	3.114354&	3.102564&	3.113299&	7.627779&	9.583552\\
Mixed Set&	2.773921&	2.770444&	2.76135&	2.777407&	2.760785&	2.759786&	6.307661&	7.696776\\
Disjoint Set&	3.179866&	3.168949&	3.166382&	3.206505&	3.176653&	3.188539&	5.333682&	6.108436\\
Random Set&	4.247034&	4.238412&	4.248383&	4.298506&	4.243931&	4.241595&	5.029729&	5.913611\\
}\loo
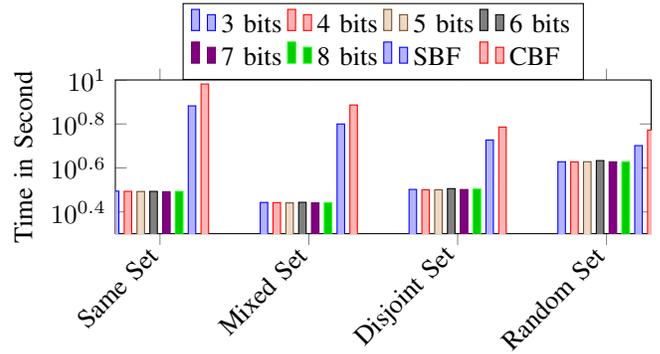
\begin{figure}[!ht]
\centering
\begin{tikzpicture}
    \begin{axis}[
            ybar,
            bar width=.1cm,
            width=0.48\textwidth,
            height=.2\textwidth,
            enlarge x limits=0.1,
            legend style={at={(0.5,1)},
                anchor=south,legend columns=4,legend cell align=left},
            symbolic x coords={Same Set, Mixed Set, Disjoint Set, Random Set},
            xtick=data,
             x tick label style={rotate=45,anchor=east},
            nodes near coords align={vertical},
            ymin=2,ymax=10,
            ymode=log,
            ylabel={Time in Second},
        ]
        \addplot table[x=interval,y=three]{\loo};
        \addplot table[x=interval,y=four]{\loo};
        \addplot table[x=interval,y=five]{\loo};
        \addplot table[x=interval,y=six]{\loo};
        \addplot table[x=interval,y=seven]{\loo};
        \addplot table[x=interval,y=eight]{\loo};
        \addplot table[x=interval,y=sbf]{\loo};
        \addplot table[x=interval,y=cbf]{\loo};
        \legend{3 bits,4 bits, 5 bits, 6 bits, 7 bits, 8 bits, SBF, CBF}
    \end{axis}
\end{tikzpicture}
\caption{Comparison of 20M items lookup time by countBF using 3 bits, 4 bits, 5 bits, 6 bits, 7 bits, and 8 bits counters with SBF and CBF. Lower is better.}
\label{look2}
\end{figure}

\pgfplotstableread[row sep=\\,col sep=&]{
interval& three&	four&	five&	six&	seven&	eight&  sbf&  cbf\\
Same Set&	4.701714&	4.721672&	4.812601&	4.688024&	4.68865&	4.695876&	11.79223&	15.120955\\
Mixed Set&	4.271157&	4.150315&	4.20941&	4.145324&	4.146619&	4.141212&	9.322156&	11.617303\\
Disjoint Set&	4.889648&	4.893726&	5.238477&	4.922776&	4.890376&	4.89394&	8.141635&	8.956921\\
Random Set&	6.837614&	6.881893&	6.882104&	6.960086&	6.833397&	6.828258&	8.016901&	8.827946\\
}\look
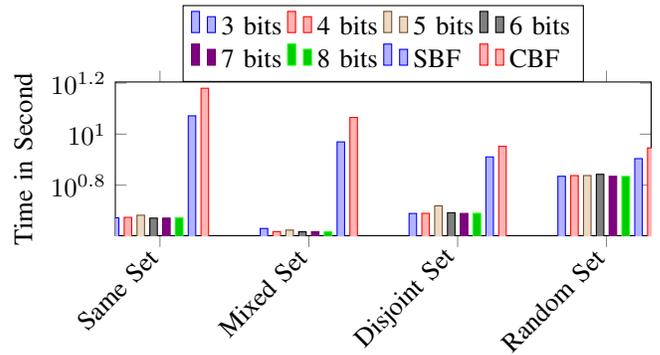
\begin{figure}[!ht]
\centering
\begin{tikzpicture}
    \begin{axis}[
            ybar,
            bar width=.1cm,
            width=0.48\textwidth,
            height=.2\textwidth,
            enlarge x limits=0.1,
            legend style={at={(0.5,1)},
                anchor=south,legend columns=4,legend cell align=left},
            symbolic x coords={Same Set, Mixed Set, Disjoint Set, Random Set},
            xtick=data,
             x tick label style={rotate=45,anchor=east},
            nodes near coords align={vertical},
            ymin=4,ymax=16,
            ymode=log,
            ylabel={Time in Second},
        ]
        \addplot table[x=interval,y=three]{\look};
        \addplot table[x=interval,y=four]{\look};
        \addplot table[x=interval,y=five]{\look};
        \addplot table[x=interval,y=six]{\look};
        \addplot table[x=interval,y=seven]{\look};
        \addplot table[x=interval,y=eight]{\look};
        \addplot table[x=interval,y=sbf]{\look};
        \addplot table[x=interval,y=cbf]{\look};
        \legend{3 bits,4 bits, 5 bits, 6 bits, 7 bits, 8 bits, SBF, CBF}
    \end{axis}
\end{tikzpicture}
\caption{Comparison of 30M items lookup time by countBF using 3 bits, 4 bits, 5 bits, 6 bits, 7 bits, and 8 bits counters with SBF and CBF. Lower is better.}
\label{look3}
\end{figure}

\pgfplotstableread[row sep=\\,col sep=&]{
interval& three&	four&	five&	six&	seven&	eight&  sbf&  cbf\\
Same Set&	6.293729&	6.294769&	6.531742&	6.270722&	6.278965&	6.285266&	15.745596&	23.890921\\
Mixed Set&	5.531775&	5.579866&	5.532112&	5.528269&	5.524324&	5.625707&	12.820064&	15.857062\\
Disjoint Set&	6.595209&	6.605372&	7.022292&	6.622624&	6.642423&	6.667282&	10.743181&	12.510179\\
Random Set&	9.50504&	9.577646&	9.81403&	9.496845&	9.558259&	9.476981&	10.542516&	12.875963\\
}\looku
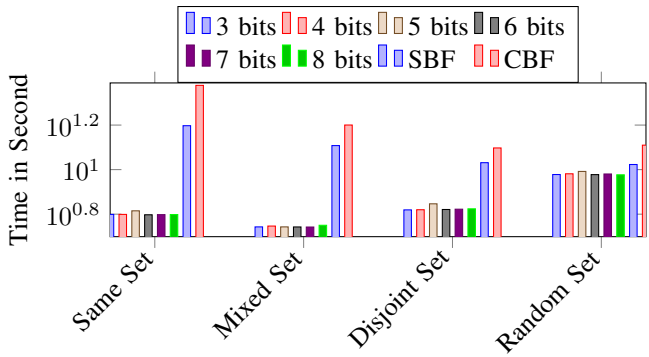
\begin{figure}[!ht]
\centering
\begin{tikzpicture}
    \begin{axis}[
            ybar,
            bar width=.1cm,
            width=0.48\textwidth,
            height=.2\textwidth,
            enlarge x limits=0.1,
            legend style={at={(0.5,1)},
                anchor=south,legend columns=4,legend cell align=left},
            symbolic x coords={Same Set, Mixed Set, Disjoint Set, Random Set},
            xtick=data,
             x tick label style={rotate=45,anchor=east},
            nodes near coords align={vertical},
            ymin=5,ymax=24.5,
            ymode=log,
            ylabel={Time in Second},
        ]
        \addplot table[x=interval,y=three]{\looku};
        \addplot table[x=interval,y=four]{\looku};
        \addplot table[x=interval,y=five]{\looku};
        \addplot table[x=interval,y=six]{\looku};
        \addplot table[x=interval,y=seven]{\looku};
        \addplot table[x=interval,y=eight]{\looku};
        \addplot table[x=interval,y=sbf]{\looku};
        \addplot table[x=interval,y=cbf]{\looku};
        \legend{3 bits,4 bits, 5 bits, 6 bits, 7 bits, 8 bits, SBF, CBF}
    \end{axis}
\end{tikzpicture}
\caption{Comparison of 40M items lookup time by countBF using 3 bits, 4 bits, 5 bits, 6 bits, 7 bits, and 8 bits counters with SBF and CBF. Lower is better.}
\label{look4}
\end{figure}

\pgfplotstableread[row sep=\\,col sep=&]{
interval& three&	four&	five&	six&	seven&	eight&   sbf&   cbf\\
Same Set&	7.887211&	8.21619&	8.000714&	7.862594&	7.871445&	8.015318&	19.591754&	24.531486\\
Mixed Set&	6.925461&	7.42123&	6.910845&	6.907221&	6.921344&	6.934987&	15.967332&	18.804615\\
Disjoint Set&	8.292553&	8.691627&	8.322802&	8.329029&	8.338787&	8.330786&	14.163228&	15.189417\\
Random Set&	12.123269&	12.133173&	12.101659&	12.158951&	12.110405&	12.080527&	13.711169&	14.877157\\
}\lookup
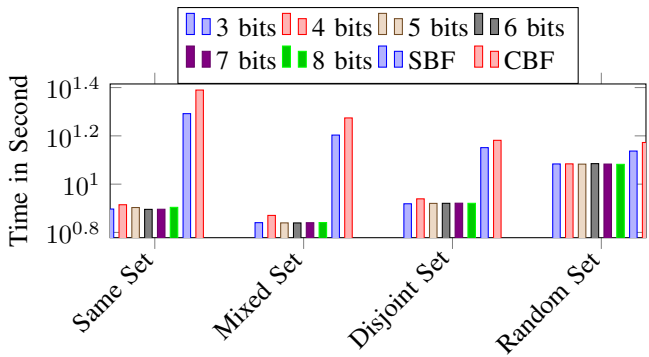
\begin{figure}[!ht]
\centering
\begin{tikzpicture}
    \begin{axis}[
            ybar,
            bar width=.1cm,
            width=0.48\textwidth,
            height=.2\textwidth,
            enlarge x limits=0.1,
            legend style={at={(0.5,1)},
                anchor=south,legend columns=4,legend cell align=left},
            symbolic x coords={Same Set, Mixed Set, Disjoint Set, Random Set},
            xtick=data,
             x tick label style={rotate=45,anchor=east},
            nodes near coords align={vertical},
            ymin=6,ymax=26,
            ymode=log,
            ylabel={Time in Second},
        ]
        \addplot table[x=interval,y=three]{\lookup};
        \addplot table[x=interval,y=four]{\lookup};
        \addplot table[x=interval,y=five]{\lookup};
        \addplot table[x=interval,y=six]{\lookup};
        \addplot table[x=interval,y=seven]{\lookup};
        \addplot table[x=interval,y=eight]{\lookup};
        \addplot table[x=interval,y=sbf]{\lookup};
        \addplot table[x=interval,y=cbf]{\lookup};
        \legend{3 bits,4 bits, 5 bits, 6 bits, 7 bits, 8 bits, SBF, CBF}
    \end{axis}
\end{tikzpicture}
\caption{Comparison of 50M items' lookup time by countBF using 3 bits, 4 bits, 5 bits, 6 bits, 7 bits, and 8 bits counters with SBF and CBF. Lower is better.}
\label{look5}
\end{figure}

Figures \ref{look1}, \ref{look2},\ref{look3}, \ref{look4}, and \ref{look5} demonstrates the lookup operations of 10M, 20M, 30M, 40M, and 50M items respectively by countBF, SBF and CBF. countBF is faster than SBF and CBF in all sized query with all test cases. On an average, countBF with 3 bits, 7 bits, 7 bits, 6 bits, and 3 bits counters are the fastest in 10M, 20M, 30M, 40M, and 50M items lookup, respectively. It is observed that countBF with 8 bits counter exhibits the worst performance overall. For any test case, countBF with 4 bits, 8 bits, 8 bits, 7 bits, and 6 bits are the fastest in the insertion of 10M, 20M, 30M, 40M, and 50M dataset, respectively. Overall, countBF with 7 bits is the fastest filter. countBF outperforms SBF and CBF by 50.73\% and	55.68\%,  45.33\% and	54.67\%, 44.84\% and 53.82\%, 43.99\% and	57.14\%, and 44.46\% and 52.01\% in 10M, 20M, 30M, 40M, and 50M items' lookup respectively. Overall, our proposed system has a higher lookup performance rate than SBF and CBF. 

\pgfplotstableread[row sep=\\,col sep=&]{
interval& three&	four&	five&	six&	seven&	eight & sbf & cbf\\
Mixed Set&	0.000010&	0.000010&	0.000010&	0.000030&	0.000020&	0.000010&	0.001028&	0.001027\\
Disjoint Set&	0.000000&	0.000001&	0.000005&	0.000009&	0.000011&	0.000014&	0.001012&	0.000996\\
Random Set&	0.000010&	0.000010&	0.000010&	0.000030&	0.000018&	0.000010&	0.000989&	0.000998\\
}\f
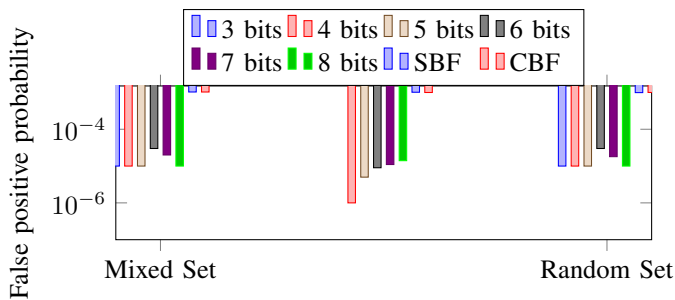
\begin{figure}[!ht]
\centering
\begin{tikzpicture}
    \begin{axis}[
            ybar,
            bar width=.1cm,
            width=0.48\textwidth,
            height=.2\textwidth,
            enlarge x limits=0.1,
            legend style={at={(0.5,1)},
                anchor=south,legend columns=4,legend cell align=left},
            symbolic x coords={Mixed Set, Disjoint Set, Random Set},
            xtick=data,
             x tick label style={anchor=north},
            nodes near coords align={vertical},
            ymin=0.0000001,ymax=0.0015,
            ymode=log,
            ylabel={False positive probability},
        ]
        \addplot table[x=interval,y=three]{\f};
        \addplot table[x=interval,y=four]{\f};
        \addplot table[x=interval,y=five]{\f};
        \addplot table[x=interval,y=six]{\f};
        \addplot table[x=interval,y=seven]{\f};
        \addplot table[x=interval,y=eight]{\f};
        \addplot table[x=interval,y=sbf]{\f};
        \addplot table[x=interval,y=cbf]{\f};
        \legend{3 bits,4 bits, 5 bits, 6 bits, 7 bits, 8 bits, SBF, CBF}
    \end{axis}
\end{tikzpicture}
\caption{Comparison of FPP in 10M lookup of countBF using 3 bits, 4 bits, 5 bits, 6 bits, 7 bits, and 8 bits counters with SBF and CBF in desired FPP of 0.001 setting. Lower is better.}
\label{fpp1}
\end{figure}

\pgfplotstableread[row sep=\\,col sep=&]{
interval& three&	four&	five&	six&	seven&	eight&  sbf&  cbf\\
Mixed Set&	0.000010&	0.000015&	0.000005&	0.000010&	0.000035&	0.000025&	0.001009&	0.000996\\
Disjoint Set&	0.000001&	0.000003&	0.000004&	0.000007&	0.000010&	0.000016&	0.001002&	0.000999\\
Random Set&	0.000009&	0.000015&	0.000005&	0.000010&	0.000035&	0.000025&	0.00098&	0.001002\\
}\fp
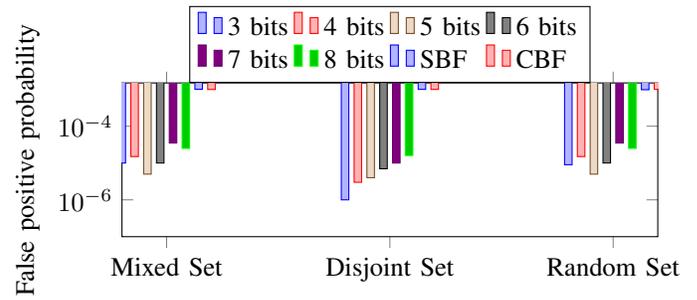
\begin{figure}[!ht]
\centering
\begin{tikzpicture}
    \begin{axis}[
            ybar,
            bar width=.1cm,
            width=0.48\textwidth,
            height=.2\textwidth,
            enlarge x limits=0.1,
            legend style={at={(0.5,1)},
                anchor=south,legend columns=4,legend cell align=left},
            symbolic x coords={Mixed Set, Disjoint Set, Random Set},
            xtick=data,
             x tick label style={anchor=north},
            nodes near coords align={vertical},
            ymin=0.0000001,ymax=0.0015,
            ymode=log,
            ylabel={False positive probability},
        ]
        \addplot table[x=interval,y=three]{\fp};
        \addplot table[x=interval,y=four]{\fp};
        \addplot table[x=interval,y=five]{\fp};
        \addplot table[x=interval,y=six]{\fp};
        \addplot table[x=interval,y=seven]{\fp};
        \addplot table[x=interval,y=eight]{\fp};
        \addplot table[x=interval,y=sbf]{\fp};
        \addplot table[x=interval,y=cbf]{\fp};
        \legend{3 bits,4 bits, 5 bits, 6 bits, 7 bits, 8 bits, SBF, CBF}
    \end{axis}
\end{tikzpicture}
\caption{Comparison of FPP in 20M lookup of countBF using 3 bits, 4 bits, 5 bits, 6 bits, 7 bits, and 8 bits counters with SBF and CBF in desired FPP of 0.001 setting. Lower is better.}
\label{fpp2}
\end{figure}

\pgfplotstableread[row sep=\\,col sep=&]{
interval& three&	four&	five&	six&	seven&	eight&  sbf&  cbf\\
Mixed Set&	0.000007&	0.000003&	0.000010&	0.000010&	0.000023&	0.000023&	0.000998&	0.000999\\
Disjoint Set&	0.000001&	0.000003&	0.000005&	0.000008&	0.000010&	0.000017&	0.001003&	0.000996\\
Random Set&	0.000007&	0.000004&	0.000011&	0.000009&	0.000023&	0.000024&	0.000981&	0.000976\\
}\fpp
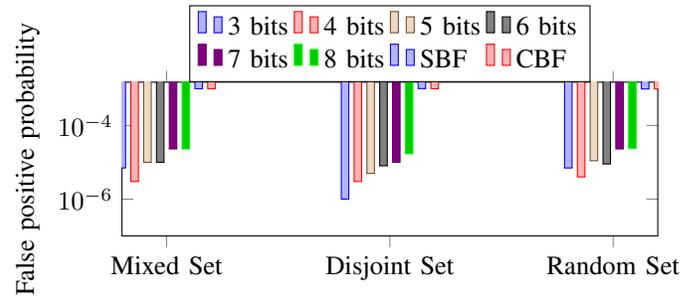
\begin{figure}[!ht]
\centering
\begin{tikzpicture}
    \begin{axis}[
            ybar,
            bar width=.1cm,
            width=0.48\textwidth,
            height=.2\textwidth,
            enlarge x limits=0.1,
            legend style={at={(0.5,1)},
                anchor=south,legend columns=4,legend cell align=left},
            symbolic x coords={Mixed Set, Disjoint Set, Random Set},
            xtick=data,
             x tick label style={anchor=north},
            nodes near coords align={vertical},
            ymin=0.0000001,ymax=0.0015,
            ymode=log,
            ylabel={False positive probability},
        ]
        \addplot table[x=interval,y=three]{\fpp};
        \addplot table[x=interval,y=four]{\fpp};
        \addplot table[x=interval,y=five]{\fpp};
        \addplot table[x=interval,y=six]{\fpp};
        \addplot table[x=interval,y=seven]{\fpp};
        \addplot table[x=interval,y=eight]{\fpp};
        \addplot table[x=interval,y=sbf]{\fpp};
        \addplot table[x=interval,y=cbf]{\fpp};
        \legend{3 bits,4 bits, 5 bits, 6 bits, 7 bits, 8 bits, SBF, CBF}
    \end{axis}
\end{tikzpicture}
\caption{Comparison of FPP of 30M lookup in countBF using 3 bits, 4 bits, 5 bits, 6 bits, 7 bits, and 8 bits counters with SBF and CBF in desired FPP of 0.001 setting. Lower is better.}
\label{fpp3}
\end{figure}

\pgfplotstableread[row sep=\\,col sep=&]{
interval& three&	four&	five&	six&	seven&	eight&  sbf&  cbf\\
Mixed Set&	0.000008&	0.000003&	0.000005&	0.000020&	0.000008&	0.000023&	0.001019&	0.001001\\
Disjoint Set&	0.000001&	0.000001&	0.000005&	0.000008&	0.000011&	0.000015&	0.001022&	0.001003\\
Random Set&	0.000008&	0.000003&	0.000005&	0.000020&	0.000007&	0.000023&	0.001&	0.000977\\
}\fppp
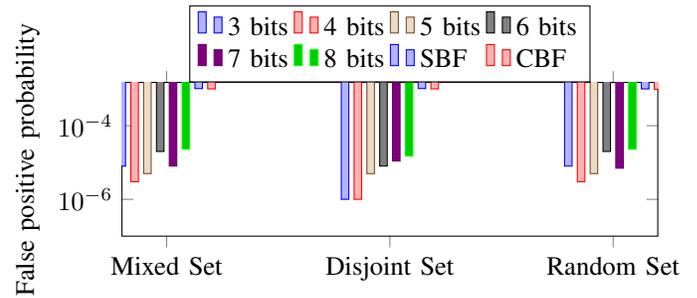
\begin{figure}[!ht]
\centering
\begin{tikzpicture}
    \begin{axis}[
            ybar,
            bar width=.1cm,
            width=0.48\textwidth,
            height=.2\textwidth,
            enlarge x limits=0.1,
            legend style={at={(0.5,1)},
                anchor=south,legend columns=4,legend cell align=left},
            symbolic x coords={Mixed Set, Disjoint Set, Random Set},
            xtick=data,
             x tick label style={anchor=north},
            nodes near coords align={vertical},
            ymin=0.0000001,ymax=0.0015,
            ymode=log,
            ylabel={False positive probability},
        ]
        \addplot table[x=interval,y=three]{\fppp};
        \addplot table[x=interval,y=four]{\fppp};
        \addplot table[x=interval,y=five]{\fppp};
        \addplot table[x=interval,y=six]{\fppp};
        \addplot table[x=interval,y=seven]{\fppp};
        \addplot table[x=interval,y=eight]{\fppp};
        \addplot table[x=interval,y=sbf]{\fppp};
        \addplot table[x=interval,y=cbf]{\fppp};
        \legend{3 bits,4 bits, 5 bits, 6 bits, 7 bits, 8 bits, SBF, CBF}
    \end{axis}
\end{tikzpicture}
\caption{Comparison of FPP of 40M lookup in countBF using 3 bits, 4 bits, 5 bits, 6 bits, 7 bits, and 8 bits counters with SBF and CBF in desired FPP of 0.001 setting. Lower is better.}
\label{fpp4}
\end{figure}

\pgfplotstableread[row sep=\\,col sep=&]{
interval& three&	four&	five&	six&	seven&	eight&  sbf&   cbf\\
Mixed Set&	0.000006&	0.000010&	0.000008&	0.000012&	0.000018&	0.000014&	0.001013&	0.000994\\
Disjoint Set&	0.000001&	0.000002&	0.000004&	0.000009&	0.000012&	0.000015&	0.001008&	0.000999\\
Random Set&	0.000006&	0.000010&	0.000008&	0.000013&	0.000018&	0.000014&	0.000976&	0.000986\\
}\fpppp
\begin{figure}[!ht]
\centering
\begin{tikzpicture}
    \begin{axis}[
            ybar,
            bar width=.1cm,
            width=0.48\textwidth,
            height=.2\textwidth,
            enlarge x limits=0.1,
            legend style={at={(0.5,1)},
                anchor=south,legend columns=4,legend cell align=left},
            symbolic x coords={Mixed Set, Disjoint Set, Random Set},
            xtick=data,
             x tick label style={anchor=north},
            nodes near coords align={vertical},
            ymin=0.0000001,ymax=0.0015,
            ymode=log,
            ylabel={False positive probability},
        ]
        \addplot table[x=interval,y=three]{\fpppp};
        \addplot table[x=interval,y=four]{\fpppp};
        \addplot table[x=interval,y=five]{\fpppp};
        \addplot table[x=interval,y=six]{\fpppp};
        \addplot table[x=interval,y=seven]{\fpppp};
        \addplot table[x=interval,y=eight]{\fpppp};
        \addplot table[x=interval,y=sbf]{\fpppp};
        \addplot table[x=interval,y=cbf]{\fpppp};
        \legend{3 bits,4 bits, 5 bits, 6 bits, 7 bits, 8 bits, SBF, CBF}
    \end{axis}
\end{tikzpicture}
\caption{Comparison of FPP in 50M lookup in countBF using 3 bits, 4 bits, 5 bits, 6 bits, 7 bits, and 8 bits counters with SBF and CBF in desired FPP of 0.001 setting. Lower is better.}
\label{fpp5}
\end{figure}
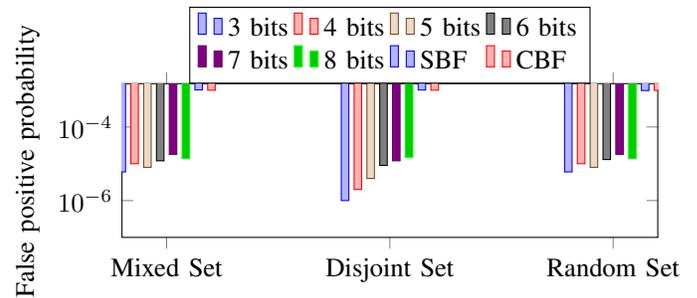

Figures \ref{fpp1}, \ref{fpp2}, \ref{fpp3}, \ref{fpp4}, and \ref{fpp5} demonstrates the false positive probability of countBF, SBF and CBF in log scale. Counting Bloom Filters are designed to deal with the false negative issue. Therefore, conventional counting Bloom Filter has an issue of trade-off between false positive probability and memory consumption. Normally, counting Bloom Filters have a higher false positive probability than other variants of the membership filter. However, countBF outperforms in the false positive probability in Mixed Set, Disjoint Set, and Random Set queries. There is no false positive probability for the Same Set query. countBF exhibits an excellent false positive probability in Disjoint Set query, and it is zero in Disjoint set for the lookup of 10M dataset, which is demonstrated in Figure \ref{fpp2}. However, the highest false positive probability of countBF is recorded as 0.000035. countBF with 7 bits counter exhibits the highest false positive probability in 20M dataset lookup. On an average, the lowest and highest false positive probability of countBF is $0.000006$ and $0.000022$, respectively. Overall, the false positive probability of countBF is $0.000013$. On an average, the false positive probability of SBF and CBF are 0.001004533333333 and 0.000999133333333, where CBF exhibits a lower false positive probability. It is possible due to the higher memory footprint of CBF than SBF. Thus, our proposed system exhibits the lowest false positive probability, whereas SBF and CBF exhibit equivalent false positive probability. Both SBF and CBF are configured to the desired false positive probability as 0.001, and therefore, their false positive probability is equivalent.


\pgfplotstableread[row sep=\\,col sep=&]{
interval& three&	four&	five&	six&	seven&	eight&  sbf&  cbf\\
Same Set&	100&	100&	100&	100&	100&	100&	100&	100\\
Mixed Set&	99.999&	99.999&	99.999&	99.997&	99.998&	99.999&	99.8972&	99.8973\\
Disjoint Set&	99.99995&	99.99994&	99.99955&	99.99912&	99.99887&	99.99858&	99.8988&	99.9004\\
Random Set&	99.99903&	99.99903&	99.99903&	99.99703&	99.99818&	99.99903&	99.9011&	99.9002\\
}\a
\begin{figure}[!ht]
\centering
\begin{tikzpicture}
    \begin{axis}[
            ybar,
            bar width=.1cm,
            width=0.48\textwidth,
            height=.2\textwidth,
            enlarge x limits=0.1,
            legend style={at={(0.5,1)},
                anchor=south,legend columns=4,legend cell align=left},
            symbolic x coords={Same Set,Mixed Set, Disjoint Set, Random Set},
            xtick=data,
             x tick label style={anchor=north},
            nodes near coords align={vertical},
            ymin=99.89,ymax=100.01,
            ylabel={Percentage},
        ]
        \addplot table[x=interval,y=three]{\a};
        \addplot table[x=interval,y=four]{\a};
        \addplot table[x=interval,y=five]{\a};
        \addplot table[x=interval,y=six]{\a};
        \addplot table[x=interval,y=seven]{\a};
        \addplot table[x=interval,y=eight]{\a};
        \addplot table[x=interval,y=sbf]{\a};
        \addplot table[x=interval,y=cbf]{\a};
        \legend{3 bits,4 bits, 5 bits, 6 bits, 7 bits, 8 bits, SBF, CBF}
    \end{axis}
\end{tikzpicture}
\caption{Comparison of accuracy in 10M lookup of countBF using 3 bits, 4 bits, 5 bits, 6 bits, 7 bits, and 8 bits counters with SBF and CBF in desired FPP of 0.001 setting. Higher is better.}
\label{acc1}
\end{figure}
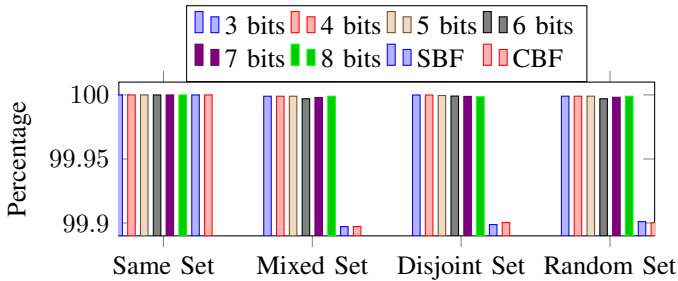

\pgfplotstableread[row sep=\\,col sep=&]{
interval& three&	four&	five&	six&	seven&	eight&  sbf&  cbf\\
Same Set&	100&	100&	100&	100&	100&	100&	100&	100\\
Mixed Set&	99.999&	99.9985&	99.9995&	99.999&	99.9965&	99.9975&	99.8991&	99.9004\\
Disjoint Set&	99.999915&	99.99973&	99.99962&	99.999335&	99.99897&	99.998365&	99.8998&	99.9001\\
Random Set&	99.99913&	99.998515&	99.99954&	99.99904&	99.996545&	99.997525&	99.902&	99.8998\\
}\ac
\begin{figure}[!ht]
\centering
\begin{tikzpicture}
    \begin{axis}[
            ybar,
            bar width=.1cm,
            width=0.48\textwidth,
            height=.2\textwidth,
            enlarge x limits=0.1,
            legend style={at={(0.5,1)},
                anchor=south,legend columns=4,legend cell align=left},
            symbolic x coords={Same Set,Mixed Set, Disjoint Set, Random Set},
            xtick=data,
             x tick label style={anchor=north},
            nodes near coords align={vertical},
            ymin=99.8,ymax=100.01,
            ylabel={Percentage},
        ]
        \addplot table[x=interval,y=three]{\ac};
        \addplot table[x=interval,y=four]{\ac};
        \addplot table[x=interval,y=five]{\ac};
        \addplot table[x=interval,y=six]{\ac};
        \addplot table[x=interval,y=seven]{\ac};
        \addplot table[x=interval,y=eight]{\ac};
        \addplot table[x=interval,y=sbf]{\ac};
        \addplot table[x=interval,y=cbf]{\ac};
        \legend{3 bits,4 bits, 5 bits, 6 bits, 7 bits, 8 bits, SBF, CBF}
    \end{axis}
\end{tikzpicture}
\caption{Comparison of accuracy in 20M lookup of countBF using 3 bits, 4 bits, 5 bits, 6 bits, 7 bits, and 8 bits counters with SBF and CBF in desired FPP of 0.001 setting. Higher is better.}
\label{acc2}
\end{figure}
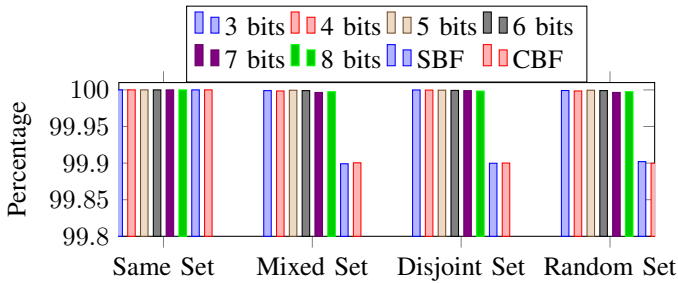

\pgfplotstableread[row sep=\\,col sep=&]{
interval& three&	four&	five&	six&	seven&	eight&  sbf&  cbf\\
Same Set&	100&	100&	100&	100&	100&	100&	100&	100\\
Mixed Set&	99.999333&	99.999667&	99.999&	99.999&	99.997667&	99.997667&	99.9002&	99.9001\\
Disjoint Set&	99.999897&	99.999727&	99.99954&	99.999247&	99.999003&	99.99831&	99.8997&	99.9004\\
Random Set&	99.999293&	99.999633&	99.998943&	99.99908&	99.997667&	99.99762&	99.9019&	99.9024\\
}\accc
\begin{figure}[!ht]
\centering
\begin{tikzpicture}
    \begin{axis}[
            ybar,
            bar width=.1cm,
            width=0.48\textwidth,
            height=.2\textwidth,
            enlarge x limits=0.1,
            legend style={at={(0.5,1)},
                anchor=south,legend columns=4,legend cell align=left},
            symbolic x coords={Same Set,Mixed Set, Disjoint Set, Random Set},
            xtick=data,
             x tick label style={anchor=north},
            nodes near coords align={vertical},
            ymin=99.8,ymax=100.01,
            ylabel={Percentage},
        ]
        \addplot table[x=interval,y=three]{\accc};
        \addplot table[x=interval,y=four]{\accc};
        \addplot table[x=interval,y=five]{\accc};
        \addplot table[x=interval,y=six]{\accc};
        \addplot table[x=interval,y=seven]{\accc};
        \addplot table[x=interval,y=eight]{\accc};
        \addplot table[x=interval,y=sbf]{\accc};
        \addplot table[x=interval,y=cbf]{\accc};
        \legend{3 bits,4 bits, 5 bits, 6 bits, 7 bits, 8 bits, SBF, CBF}
    \end{axis}
\end{tikzpicture}
\caption{Comparison of accuracy in 30M lookup of countBF using 3 bits, 4 bits, 5 bits, 6 bits, 7 bits, and 8 bits counters with SBF and CBF in desired FPP of 0.001 setting. Higher is better.}
\label{acc3}
\end{figure}
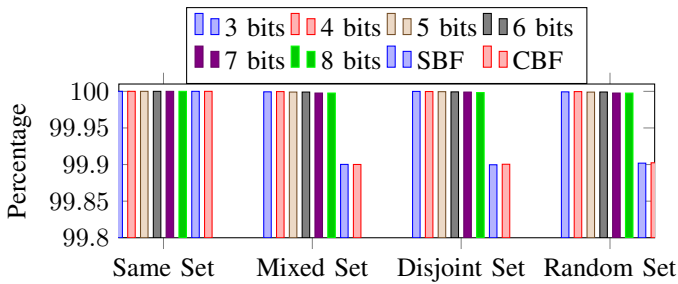

\pgfplotstableread[row sep=\\,col sep=&]{
interval& three&	four&	five&	six&	seven&	eight&  sbf&  cbf\\
Same Set&	100&	100&	100&	100&	100&	100&	100&	100\\
Mixed Set&	99.99925&	99.99975&	99.9995&	99.998&	99.99925&	99.99775&	99.8981&	99.8999\\
Disjoint Set&	99.999923&	99.999877&	99.999547&	99.999198&	99.998895&	99.99853&	99.8978&	99.8997\\
Random Set&	99.999233&	99.999728&	99.999493&	99.998015&	99.999305&	99.997715&	99.9&	99.9023\\
}\acccc
\begin{figure}[!ht]
\centering
\begin{tikzpicture}
    \begin{axis}[
            ybar,
            bar width=.1cm,
            width=0.48\textwidth,
            height=.2\textwidth,
            enlarge x limits=0.1,
            legend style={at={(0.5,1)},
                anchor=south,legend columns=4,legend cell align=left},
            symbolic x coords={Same Set,Mixed Set, Disjoint Set, Random Set},
            xtick=data,
             x tick label style={anchor=north},
            nodes near coords align={vertical},
            ymin=99.8,ymax=100.01,
            ylabel={Percentage},
        ]
        \addplot table[x=interval,y=three]{\acccc};
        \addplot table[x=interval,y=four]{\acccc};
        \addplot table[x=interval,y=five]{\acccc};
        \addplot table[x=interval,y=six]{\acccc};
        \addplot table[x=interval,y=seven]{\acccc};
        \addplot table[x=interval,y=eight]{\acccc};
        \addplot table[x=interval,y=sbf]{\acccc};
        \addplot table[x=interval,y=cbf]{\acccc};
        \legend{3 bits,4 bits, 5 bits, 6 bits, 7 bits, 8 bits, SBF, CBF}
    \end{axis}
\end{tikzpicture}
\caption{Comparison of accuracy in 40M lookup of countBF using 3 bits, 4 bits, 5 bits, 6 bits, 7 bits, and 8 bits counters with SBF and CBF in desired FPP of 0.001 setting. Higher is better.}
\label{acc4}
\end{figure}
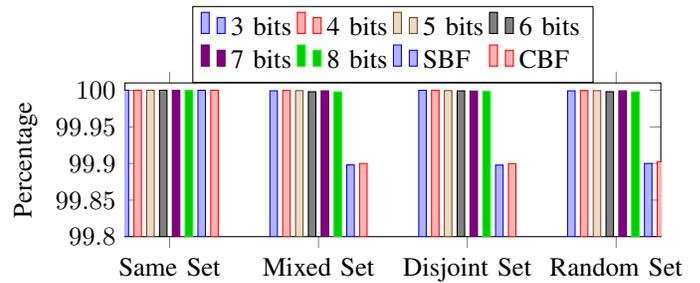

\pgfplotstableread[row sep=\\,col sep=&]{
interval& three&	four&	five&	six&	seven&	eight&  sbf& cbf\\
Same Set&	100&	100&	100&	100&	100&	100&	100&	100\\
Mixed Set&	99.9994&	99.999&	99.9992&	99.9988&	99.9982&	99.9986&	99.8987&	99.9006\\
Disjoint Set&	99.99992&	99.999822&	99.999608&	99.999058&	99.998766&	99.998464&	99.8992&	99.9001\\
Random Set&	99.999384&	99.998972&	99.99921&	99.998728&	99.99817&	99.998552&	99.9024&	99.9014\\
}\accccc
\begin{figure}[!ht]
\centering
\begin{tikzpicture}
    \begin{axis}[
            ybar,
            bar width=.1cm,
            width=0.48\textwidth,
            height=.2\textwidth,
            enlarge x limits=0.1,
            legend style={at={(0.5,1)},
                anchor=south,legend columns=4,legend cell align=left},
            symbolic x coords={Same Set,Mixed Set, Disjoint Set, Random Set},
            xtick=data,
             x tick label style={anchor=north},
            nodes near coords align={vertical},
            ymin=99.8,ymax=100.01,
            ylabel={Percentage},
        ]
        \addplot table[x=interval,y=three]{\accccc};
        \addplot table[x=interval,y=four]{\accccc};
        \addplot table[x=interval,y=five]{\accccc};
        \addplot table[x=interval,y=six]{\accccc};
        \addplot table[x=interval,y=seven]{\accccc};
        \addplot table[x=interval,y=eight]{\accccc};
        \addplot table[x=interval,y=sbf]{\accccc};
        \addplot table[x=interval,y=cbf]{\accccc};
        \legend{3 bits,4 bits, 5 bits, 6 bits, 7 bits, 8 bits, SBF, CBF}
    \end{axis}
\end{tikzpicture}
\caption{Comparison of accuracy in 50M lookup of countBF using 3 bits, 4 bits, 5 bits, 6 bits, 7 bits, and 8 bits counters with SBF and CBF in desired FPP of 0.001 setting. Higher is better.}
\label{acc5}
\end{figure}
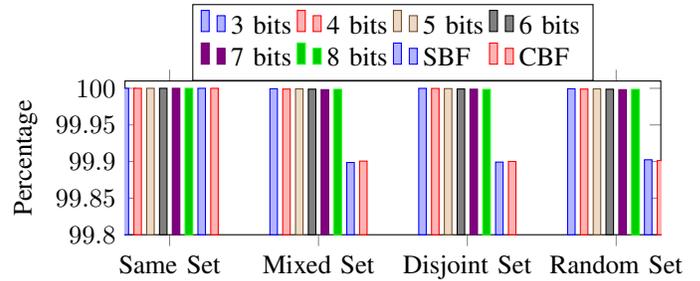

Figures \ref{acc1}, \ref{acc2}, \ref{acc3}, \ref{acc4}, and \ref{acc5} show the accuracy of countBF, SBF and CBF. The accuracy of SBF and CBF does not differ more, and both are equivalent due to the desired false positive setting to 0.001. However, the accuracy of countBF outperforms SBF and CBF in all sized dataset with all test cases. The highest accuracy is 100\% in Disjoint Set dataset of 20M lookup, and it is countBF with 3 bits counter, which is demonstrated in Figure \ref{acc2}. On an average, the lowest and highest accuracy of countBF are 99.999897\%, and 99.99995\% respectively. Overall, countBF accuracy is $99.999921\%$. The overall accuracy of SBF and CBF are 99.8997\% and 99.90\%, respectively. 

\pgfplotstableread[row sep=\\,col sep=&]{
interval& countBF&  sbf& cbf\\
10M&	8.799034&	17.13942&	68.557697\\
20M&	17.509224&	34.27884&	137.115379\\
30M&	26.365181&	51.41826&	205.673057\\
40M&	34.776222&	68.55768&	274.230739\\
50M&	43.106789&	85.6971&	342.788415909\\
}\mem
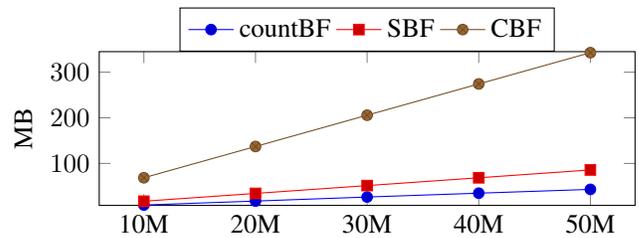
\begin{figure}[!ht]
\centering
\begin{tikzpicture}
    \begin{axis}[
            width=0.48\textwidth,
            height=.2\textwidth,
            enlarge x limits=0.1,
            legend style={at={(0.5,1)},
                anchor=south,legend columns=4,legend cell align=left},
            symbolic x coords={10M,20M,30M,40M,50M},
            xtick=data,
             x tick label style={anchor=north},
            nodes near coords align={vertical},
            ymin=8,ymax=345,
            ylabel={MB},
        ]
        \addplot table[x=interval,y=countBF]{\mem};
        \addplot table[x=interval,y=sbf]{\mem};
        \addplot table[x=interval,y=cbf]{\mem};
        \legend{countBF, SBF, CBF}
    \end{axis}
\end{tikzpicture}
\caption{Comparison of memory consumption of countBF, SBF and CBF in desired FPP of 0.001 setting. Lower is better.}
\label{memory}
\end{figure}

As we know that the CBF has a higher false positive probability than SBF. On the contrary, the experimental results show that SBF and CBF similar false positive probability. We have indeed configured the desired false positive probability to 0.001, and therefore, the false positive probability of SBF and CBF are equivalent. However, the memory requirements are different. CBF uses higher memory than SBF. It means CBF has a higher false positive probability than SBF. We adjust the memory allocation to achieve the desired false positive probability. Therefore, CBF has allocated more memory to achieve the desired false positive probability. Thus, counting Bloom Filter occupies more memory to achieve the desired false positive probability. Notably, counting Bloom Filter consume more memory than other variants of Bloom Filter or membership filter to achieve certain false positive probability. On the contrary, SBF uses $1.96\times$ more memory than countBF, and CBF uses $7.85\times$ more memory than countBF on an average. The lower memory footprint of countBF shows the highest accuracy and the lowest false positive probability. On an average, countBF, SBF and CBF consume memory of 26.11 MB, 51.42 MB, and 205.67 MB for all datasets, respectively. The memory, false positive probability, and accuracy are the key decisive factor of the Bloom Filter. However, countBF also faster than SBF and CBF. But there are much faster membership filters available, however, countBF is more accurate than any other counting variant of Bloom Filters.

\pgfplotstableread[row sep=\\,col sep=&]{
interval& countBF&  sbf& cbf\\
10M& 7.3811647004672&	14.377587572736&	57.5103645515776\\
20M& 7.3439008260096&	14.377587572736&	57.5103582601216\\
30M& 7.37223894193493&	14.377587572736&	57.5103550444885\\
40M& 7.2931023519744&	14.377587572736&	57.5103542755328\\
50M& 7.23211910119424&	14.377587572736&	57.5103529600313\\
}\bpi
\begin{figure}[!ht]
\centering
\begin{tikzpicture}
    \begin{axis}[
            ybar,
            bar width=.1cm,
            width=0.48\textwidth,
            height=.16\textwidth,
            enlarge x limits=0.1,
            legend style={at={(0.5,1)},
                anchor=south,legend columns=4,legend cell align=left},
            symbolic x coords={10M,20M,30M,40M,50M},
            xtick=data,
             x tick label style={anchor=north},
            nodes near coords align={vertical},
            ymin=5,ymax=60,
            ylabel={MB},
        ]
        \addplot table[x=interval,y=countBF]{\bpi};
        \addplot table[x=interval,y=sbf]{\bpi};
        \addplot table[x=interval,y=cbf]{\bpi};
        \legend{countBF, SBF, CBF}
    \end{axis}
\end{tikzpicture}
\caption{Comparison of bits per item of countBF, SBF and CBF. Lower is better.}
\label{bpi}
\end{figure}
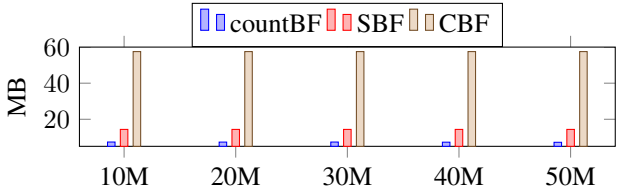

Figure \ref{bpi} demonstrates the bits per item of the countBF, SBF and CBF. countBF uses the lowest bits per item, and CBF uses the highest bits per item. countBF, SBF, and CBF use 7.32 bits, 14.38 bits, and 57.51 bits per item on average, respectively. Our proposed counting Bloom Filter uses the lowest bits per item and provides a lower false positive probability.

\pgfplotstableread[row sep=\\,col sep=&]{
interval& Waste \\
Three & 1 \\
Four & 0 \\
Five & 4 \\
Six & 4 \\
Seven & 1\\
Eight & 0\\
}\w
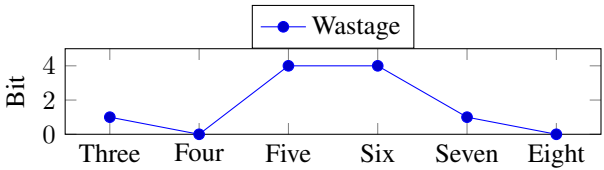
\begin{figure}[!ht]
\centering
\begin{tikzpicture}
    \begin{axis}[
            width=0.48\textwidth,
            height=.15\textwidth,
            enlarge x limits=0.1,
            legend style={at={(0.5,1)},
                anchor=south,legend columns=4,legend cell align=left},
            symbolic x coords={Three,Four,Five,Six,Seven,Eight},
            xtick=data,
             x tick label style={anchor=north},
            nodes near coords align={vertical},
            ymin=0,ymax=5,
            ylabel={Bit},
        ]
        \addplot table[x=interval,y=Waste]{\w};
        \legend{Wastage}
    \end{axis}
\end{tikzpicture}
\caption{Comparison of memory wastage in bits per memory cell of countBF with 3 bits, 4 bits, 5 bits, 6 bits, 7 bits, and 8 bits counters. Assuming, these countBF occupies 64-bits memory in each cell, and therefore, it shows the total number of unused bits (allocated but not used). Lower is better.}
\label{mem}
\end{figure}

Figure \ref{mem} depicts each cell's memory wastage in countBF with 3 bits, 4 bits, 5 bits, 6 bits, 7 bits, and 8 bits counters. countBF is a 2DBF structure that comprises many cells, and the size of each cell depends on the data type allocated for the 2D array. We use \textbf{unsigned long int} 2D array where each cell occupies 64 bits memory for our experimentation. Thus, each cell's memory wastage in countBF with 3 bits, 4 bits, 5 bits, 6 bits, 7 bits, and 8 bits counters are 1 bit, 0 bits, 4 bits, 4 bits, 1 bit and 0 bits, respectively. Figure \ref{mem} shows the allocated bits but not used, for instance, a cell size is 64 bits, and a counter size is 5 bits, then $\frac{64}{5}$ and remainder is 4. Thus, four bits are left from each cell.

\section{Analysis}
\label{Ana}
countBF is derived from 2DBF and conventional Bloom Filter. Therefore, we derive the memory requirements from conventional Bloom Filter. Let, $m$ be the bit size of Bloom Filter, $n$ be the total number of input items, and $k$ be the total number of hash functions. Therefore, the probability of a particular bit is not set to 1 is $(1-\frac{1}{m})$. The probability of that particular bit is not set to 1 by $k$ hash functions is $(1-\frac{1}{m})^k$. There are $n$ input items and the probability of that particular bit is not set to 1 is $(1-\frac{1}{m})^{kn}\approx e^{-kn/m}$. The probability of the particular bit to be 1 is $(1-e^{-kn/m})$. The probability of all bits are set to 1 is $\varepsilon=(1-e^{-kn/m})^k$ where $\varepsilon$ is the desired false positive probability. Conventional Bloom Filter exhibits its best performance in $k=\frac{m}{n}ln~2$. Simplifying the desired false positive probability using the value of $k$, we get $ln\varepsilon=-\frac{m}{n}(ln~2)^2$. Therefore, the total memory requirement for desired false positive probability $\varepsilon$ with $n$ input items is $m=-\frac{n~ln\varepsilon}{(ln~2)^2}$. However, the memory requirements of countBF depends on dimensions of 2DBF. Therefore, countBF uses $k=\frac{k}{2}$ hash functions. The memory requirement is calculated as follows- we divide the total memory requirement of $m$ using $b=\frac{m}{2\times\beta}$ where $\beta$ is the memory bits occupied by each cell in countBF. Now, $q=\sqrt{b}$ to select the $q^{th}$ prime number. Let, $\mathcal{PN}_\psi$ be the array of $\psi$ prime numbers. Let, $i=$\textsc{slectPrimeNumber}($q$) select a prime number from $\mathcal{PN}_\psi$ indexed at $q$ and assigned the returned value to $i$.  The dimension of countBF is calculated as $X=\mathcal{PN}_{i+3}$ and $Y=\mathcal{PN}_{i-3}$, because the dimension of countBF cannot be same, i.e., $X\not=Y$ and these are prime numbers. Therefore, the total memory requirement is $X*Y*\beta$ bits or $\frac{X*Y*\beta}{MB}$ MB where $MB=8\times 1024\times 1024$ is a megabyte. 

We exploit the property of the hashing technique with prime numbers. Prime number reduces collision in hashing. Therefore, our proposed counting Bloom Filter uses two prime numbers to perform modulus operation and define the dimension of the 2D filter. Thus, two modulus operations using prime numbers significantly reduce the false positive probability. Therefore, countBF is able to increase its accuracy using a lower memory footprint. Moreover, we are able to reduce the number of hash functions to half. 

\subsection{Comparison of countBF with other filters}

\begin{table}[!ht]
    \centering
    \caption{Comparison among a few membership filter. Rating scale from 1-10. Rating 1 is the worst and 10 is the best. }
    \begin{tabular}{|p{1.5cm}|p{1.5cm}|p{1cm}|p{1.5cm}|p{1cm}|}
    \hline
       Filter &  False positives & Accuracy&  Memory Consumption & Query Speed \\ \hline 
       SBF \cite{Kirsch} & 7 & 7 & 6 &  6 \\ \hline
       CBF \cite{countingBF} & 4 & 4 & 3 & 5 \\ \hline
       Cuckoo Filter \cite{Cuckoo} & 4 & 4 & 5 & 9 \\ \hline
       CQF \cite{CQF} & 6 & 6 & 7 & 9 \\ \hline
       Morton Filter \cite{Morton} & 6 & 6 & 7 & 10 \\ \hline
       XOR Filter \cite{XOR} & 6 & 6 & 7 & 10 \\ \hline
       VI-CBF \cite{Rottenstreich} & 6 & 6 & 5 & 9\\ \hline
       TCBF \cite{TCBF} & 7 & 7 & 7 & 9 \\ \hline
       countBF & 10 & 10 & 10 & 8 \\ \hline
    \end{tabular}
    \label{tabc}
\end{table}

There are numerous faster membership filters available than countBF. For instance, counting Quotient Filter (CBF) \cite{CQF}, Morton Filter (MF) \cite{Morton}, XOR Filter (XF) \cite{XOR}. Morton and XOR filters are the fastest filters. However, there is a trade-off among speed, memory and false positive probability. Also, there is a compressed Bloom Filter, which compromises performance and accuracy with memory \cite{Comp}. Various membership filters are available and a few membership filters are compared in Table \ref{tabc}. Many of the membership filters do not provide high accuracy and low false positive probability with a tiny amount of memory, while countBF can do the same. CBF has the highest false positive probability and consume the highest memory \cite{countingBF}. The lookup and query performance are same for Cuckoo Filter (CF) \cite{Cuckoo} and countBF. Cuckoo Filter exhibits poor in the false positive probability.

Table \ref{tabc} demonstrates the overall comparison among SBF, CBF, CF, CQF, MF, XF, VI-CBF, TCBF and countBF using rating points from 1 to 10. Rating 1 represents the worst, and rating 10 is the best. SBF is the standard, and the overall rating is 6.5. MF and XF are the fastest, and their ratings are 10 for both. The rating of countBF and CF is the same in execution time while countBF outperforms in other features. Moreover, SBF, CF, MF, and XF are the same in false positive probability and accuracy. SBF, CF, MF, XF, VI-CBF, and TCBF have an almost similar memory footprint. countBF outperforms all other filters in false positive probability, accuracy, and memory footprint except execution times. 

\subsection{Frequency count}
 Count-Min Sketch (CMS) is a probabilistic data structure for frequency count of input stream \cite{CMS}. countBF is converted into CMS. Insertion of an item causes an increment of $k$ counters by $k$ hash functions. Counter value can be extracted by extract masks in countBF and let these counters be $\mathcal{C}_1,~\mathcal{C}_2,~\mathcal{C}_3, \ldots \mathcal{C}_k$. These counters are placed in different locations of countBF depending on the hash function. Therefore, the frequency of an item $\mathcal{K}$ in countBF can be found in Equation \eqref{eqcms1}.  
\begin{equation}\label{eqcms1}
    Count(\mathcal{K})=\min_{i=1~to~k} \mathcal{C}_i
\end{equation}
The minimum count among all counters will be the frequency of $\mathcal{K}$. This frequency count is an approximation counting. CMS requires a large number of bits in each counters and high accuracy. These capabilities have already been demonstrated experimentally and evaluated using different parameters.


\section{Related works and Discussion}
\label{Dis}
Counting Bloom Filter (CBF) is introduced on 2000 by Li \textit{et al.} \cite{countingBF}. Counting Bloom Filter become popular due to false negative free. Deletion introduces false negatives in conventional Bloom Filter \cite{ICOIN}, but deletion is the most important operation in many applications. Therefore, diverse variants of counting Bloom Filter have been introduced, and Luo \textit{et al.} \cite{Luo} reported 15 counting Bloom Filters till 24 December 2018. Recently, a few counting Bloom Filters are developed,  tandem counting Bloom Filter (TCBF) \cite{TCBF}, Einziger and Friedman \cite{Einziger}, and mergeCBF \cite{Liu}. Current literature shows that even if there are diverse counting Bloom Filters available, but are unable to reduce memory consumption per element. Even the comparatively memory consumption per item of state-of-the-art CBFs are higher than SBF. False positive and memory have a strong relationship. Higher memory is required to lower the false positive probability. Thus, we can conclude that counting Bloom Filter has a higher false positive probability than SBF \cite{Kirsch}. The key challenge is to decrease false positive probability using less memory without compromising the insertion/query/deletion operations' performance. We have already demonstrated experimentally that our proposed countBF uses less memory footprint than any state-of-the-art counting Bloom Filters. Also, countBF can increase its accuracy by reducing false positive probability using a low memory footprint. 


\section{Conclusion}
\label{Con}
This article has demonstrated our proposed counting Bloom Filter, called countBF, significantly improved over SBF and CBF. countBF can insert, query, and delete an item in $O(k)$ time complexity while preserving high accuracy, low false positive probability, low memory footprint, and fast execution time. Moreover, the properties of countBF make more room for incoming items. We have evaluated the false positive probability using various test cases experimentally. The false positive probability is lower than the SBF and CBF. Alternatively, the accuracy of countBF is higher than SBF and CBF. Therefore, countBF is able to outperform the existing Bloom Filter in terms of false positives, accuracy, memory footprint, and performance. Also, we have compared with various state-of-the-art counting Bloom Filters and it shows that our proposed counting Bloom Filter is an ideal solution. Moreover, we have demonstrated how to adapt countBF in frequency count, similar to CMS. It can also be applied in diverse domains where delete operation is a crucial part of the system.
\balance
\bibliographystyle{IEEEtran}
\bibliography{mybib}
\end{document}